\pdfoutput=1
\RequirePackage{ifpdf}
\ifpdf 
\documentclass[pdftex]{sigma}
\else
\documentclass{sigma}
\fi

\numberwithin{equation}{section}

\newtheorem{Theorem}{Theorem}[section]
\newtheorem*{Theorem*}{Theorem}

 { \theoremstyle{definition}
\newtheorem{Definition}[Theorem]{Definition}

 }

\newcommand{\mbold}[1]{\mbox{\boldmath{\ensuremath{#1}}}}

\def\beq#1\eeq{\begin{align}#1\end{align}}

\def \a {\alpha}
\def \b {\beta}
\def \e {\gamma}

\def \bh {\mbox{{\bf h}}}

\def \tv {\tilde{ V}}
\def \bomega {\mbox{{\mbold \omega}}}

\begin{document}

\allowdisplaybreaks

\renewcommand{\thefootnote}{}

\newcommand{\arXivNumber}{2401.00780}

\renewcommand{\PaperNumber}{078}

\FirstPageHeading

\ShortArticleName{Symmetries in Riemann--Cartan Geometries}

\ArticleName{Symmetries in Riemann--Cartan Geometries\footnote{This paper is a~contribution to the Special Issue on Symmetry, Invariants, and their Applications in honor of Peter J.~Olver. The~full collection is available at \href{https://www.emis.de/journals/SIGMA/Olver.html}{https://www.emis.de/journals/SIGMA/Olver.html}}}

\Author{David D. MCNUTT~$^{\rm a}$, Alan A. COLEY~$^{\rm b}$, Robert J. VAN DEN HOOGEN~$^{\rm c}$}

\AuthorNameForHeading{D.D.~McNutt, A.A.~Coley, R.J.~van~den~Hoogen}

\Address{$^{\rm a)}$~Center for Theoretical Physics, Polish Academy of Sciences, Warsaw, Poland}
\EmailD{\href{mailto:mcnuttdd@gmail.com}{mcnuttdd@gmail.com}}

\Address{$^{\rm b)}$~Department of Mathematics and Statistics, Dalhousie University,\\
\hphantom{$^{\rm b)}$}~Halifax, Nova Scotia, Canada}
\EmailD{\href{mailto:alan.coley@dal.ca}{alan.coley@dal.ca}}

\Address{$^{\rm c)}$~Department of Mathematics and Statistics, St. Francis Xavier University,\\
\hphantom{$^{\rm c)}$}~Antigonish, Nova Scotia, Canada}
\EmailD{\href{mailto:rvandenh@stfx.ca}{rvandenh@stfx.ca}}

\ArticleDates{Received January 02, 2024, in final form August 21, 2024; Published online September 01, 2024}

\Abstract{Riemann--Cartan geometries are geometries that admit non-zero curvature and torsion tensors. These geometries have been investigated as geometric frameworks for potential theories in physics including quantum gravity theories and have many important differences when compared to Riemannian geometries. One notable difference, is the number of symmetries for a Riemann--Cartan geometry is potentially smaller than the number of Killing vector fields for the metric. In this paper, we will review the investigation of symmetries in Riemann--Cartan geometries and the mathematical tools used to determine geometries that admit a given group of symmetries. As an illustration, we present new results by determining all static spherically symmetric and all stationary spherically symmetric Riemann--Cartan geometries. Furthermore, we have determined the subclasses of spherically symmetric Riemann--Cartan geometries that admit a~seven-dimensional group of symmetries.}


\Classification{53A55; 83D99; 53Z05}

\renewcommand{\thefootnote}{\arabic{footnote}}
\setcounter{footnote}{0}

\section{Introduction} \label{sec:intro}

Riemann--Cartan geometry is a generalization of Riemannian geometry where the non-metric connection associated with covariant differentiation $\nabla$ is still required to be metric compatible. Therefore, the metric ${\bf g}$ is constant under covariant differentiation $\nabla {\bf g}( {\bf X}, {\bf Y}) =0$ but $\nabla$ is no longer torsion-free. That is, the connection admits a non-zero tensor
\[
 {\bf T}({\bf X}, {\bf Y}) = \nabla_{{\bf X}} {\bf Y} - \nabla_{{\bf Y}} {\bf X}- \mathcal{L}_{{\bf X}} {\bf Y} \neq 0,
 \]
 where $\mathcal{L}_{{\bf X}}$ denotes the Lie derivative with respect to ${\bf X}$. By allowing a non-trivial torsion, the connection is no longer guaranteed to be uniquely defined in terms of the metric, unlike the Levi-Civita connection in Riemannian geometry. However, given a torsion tensor, the connection can be uniquely specified by taking the Levi-Civita connection and adding the contorsion tensor, which is constructed from the torsion tensor using index notation as \cite{Trautman:2006fp}
\[
K_{abc} = \frac12 ( T_{abc} + T_{bca} + T_{cab}).
 \]
 If the torsion tensor is unspecified, the choice of connection cannot be specified uniquely and this freedom in the connection can complicate characterizing Riemann--Cartan geometries. Despite this complication, we will denote a Riemann--Cartan geometry, as a triple $(M, {\bf g}, \bomega)$ where $M$ is the manifold, ${\bf g}$ is the metric, and $\bomega$ is the metric-compatible connection defining $\nabla$.

The motivation for studying Riemann--Cartan geometries includes torsion based theories of gravitation, such as the Einstein--Cartan theory of gravitation, where torsion is associated with the intrinsic angular momentum \cite{Hehl_vonderHeyde_Kerlick_Nester1976,Trautman:2006fp}. This association is partially motivated by the search for a quantum theory of gravity and the unification of gravity with other fundamental interactions arising in the quantum domain. Alternatively, within the class of teleparallel gravity theories, the curvature tensor vanishes and all information about the gravitational field is encoded in the torsion tensor and its covariant derivatives. In teleparallel gravity theories, the metric is subsumed by a frame basis as the primary object of study while the affine connection can be computed from some, as yet arbitrary Lorentz frame transformation contained within the spin connection plus the Weitzenbock connection associated with the frame basis \cite{Krssak_Pereira2015}.

In any theory of gravity, it is important to determine solutions that reflect our physical Universe. However, due to the freedom in the choice of the connection, generating physically meaningful solutions in Riemann--Cartan geometry is more difficult than in theories of gravity based on Riemann geometry. Furthermore, the Ricci tensor is no longer necessarily symmetric, and depending on the gravity theory, the requirement that the Ricci tensor is related to a~symmetric energy momentum tensor will impose additional conditions that must be satisfied~\cite{tsamparlis1981methods}.

Symmetries and symmetry methods are powerful tools in the analysis of complex mathematical and physical problems. Techniques based on symmetry have been used in general relativity (GR) to produce meaningful solutions with closed form expressions. An important example of this is the Schwarzschild solution, which describes a spherically symmetric mass in vacuum and provides the simplest model for a vacuum black hole solution in GR. This was accomplished by requiring that the solution has spherical symmetry constraining the metric and curvature, which contains all of the information about the gravitational field within GR.

To illustrate, we note that any continuous symmetry can be generated by an infinitesimal vector field ${\bf X}$ and if the metric ${\bf g}$ is invariant under such a symmetry then $\mathcal{L}_{{\bf X}} {\bf g} = 0$. The resulting first order linear partial differential equations are known as the Killing equations whose solutions are Killing symmetries or isometries. The study of Killing symmetries is well understood and they have been used to find a wealth of solutions in GR \cite{kramer}.

While GR is based on Riemannian geometry, where symmetry techniques have been established, in the case of alternative theories of gravity which are not based on Riemannian geometry, less is known. Symmetries and application of symmetry techniques are currently under development, particularly in teleparallel gravity \cite{Coley:2019zld, Hohmann:2021ast, Hohmann:2019nat, McNutt:2023nxmnxm, Pfeifer:2022txm}. Here, we are interested in using symmetry techniques to determine the constraints on the geometrical framework in Riemann--Cartan geometries, where many of the approaches in teleparallel gravity no longer apply.

The introduction of a non-vanishing torsion tensor introduces additional conditions for a~symmetry generator ${\bf X}$, namely,
\beq \mathcal{L}_{{\bf X}} {\bf g} = 0, \qquad \mathcal{L}_{{\bf X}} {\bf T} = 0, \label{NaiveKV} \eeq
 and these conditions can complicate the symmetry techniques used in GR. Any symmetry satisfying these conditions will be called an affine frame symmetry. For example, it has already been shown that the only Riemann--Cartan geometry that admits a maximal number of ten affine frame symmetries is Minkowski space in which the torsion vanishes \cite{Coley:2019zld}. Furthermore, any Riemann–Cartan geometry with non-zero torsion admits at most a seven-dimensional affine frame symmetry group \cite[Corollary IV.1]{Coley:2019zld}. This is proven by applying the most general definition of a symmetry within the framework of the Cartan--Karlhede algorithm and examining the linear isotropy group of the torsion tensor.

If the metric and torsion tensor is given, equation \eqref{NaiveKV} is sufficient to determine if a vector field ${\bf X}$ is an affine frame symmetry, even if the symmetry group admits non-trivial isotropy. However, if instead one wishes to determine the most general class of Riemann--Cartan geometries which admits a given Lie group as a symmetry group, equation \eqref{NaiveKV} leads to a significantly more difficult problem. However, by switching to a frame-based perspective, this problem can be answered for any symmetry group. More concretely, in metric-affine theories, the quantities in equation \eqref{NaiveKV} can instead be represented in terms of the fundamental variables, the metric~${\bf g}$, the frame basis $\{ \bh^a\}$ and the connection~$\bomega$ \cite{Hehl_McCrea_Mielke_Neeman1995}.

In this paper, we will review the current frame based symmetry techniques in Riemann--Cartan geometry; namely, the affine frame symmetry approach \cite{McNutt:2023nxmnxm} and the Cartan--Karlhede algorithm. Using these tools, we will summarize the construction of spherically symmetric Riemann--Cartan geometries, specialize to static spherically symmetric and stationary spherically symmetric Riemann--Cartan geometries and determine the subclasses of spherically symmetric Riemann--Cartan geometries that admit a seven-dimensional affine frame symmetry group.\looseness=-1

\subsection{Notation}

Coordinate indices will be denoted by $\mu, \nu, \dots$ while tangent space indices will be denoted by~${a,b,\dots}$. The spacetime coordinates will be $x^\mu$, unless specified otherwise. The frame fields are denoted as $\bh_a$ and the dual coframe one-forms are $\bh^a$. Then relative to a coordinate basis, the vielbein components are $h_a^{~\mu}$ or $h^a_{~\mu}$. The spacetime metric will be denoted as $g_{\mu \nu}$ while the Minkowski tangent space metric is $\eta_{ab}$.

To denote a local Lorentz transformation leaving $\eta_{ab}$ invariant, we will write $\Lambda_a^{~b}(x^\mu)$ and \smash{$\Lambda^a_{~b}  =  \big(\Lambda^{-1}\big)_a^{~b}$} as its inverse transformation. The connection one-form $\bomega^a_{~b}$ is defined as \linebreak ${\bomega^a_{~b}  =  \omega^a_{~bc} \bh^c}$, and the metric-compatibility condition implies that the first two lowered indices are anti-symmetric. The curvature and torsion tensors will be denoted, respectively, as $R^a_{~bcd}$ and $T^a_{~bc}$.

Relative to a frame basis $\{ \bh_a \}$ covariant derivatives with respect to a basis element will be denoted as $\nabla_a \equiv \nabla_{\bh_a}$. Covariant derivatives with respect to a metric-compatible connection will be denoted using a semi-colon, $T_{abc;e}$. Integral curves of a vector field ${\bf X}$, with parameter $\tau$, will be denoted as $\phi_{\tau} \colon M \to M$ and we will write the corresponding Lorentz frame transformation as $\phi_\tau^{*} \bh^a = \Lambda^a_{(\tau) b} \bh^b$.

Due to the variety of commutator relations used in the affine frame symmetry formalism, we will use different indices to distinguish them. The commutators of the affine frame symmetry generators $\{{\bf X}_I \}$ are $[{\bf X}_I, {\bf X}_J] = C^K_{~IJ} {\bf X}_K$. The Lie algebra representations of the isotropy group, $ \lambda_{\hat{i}} =\lambda^a_{~\hat{i} b}$ are \smash{$[\lambda_{\hat{i}}, \lambda_{\hat{j}}] = C^{\hat{k}}_{~\hat{i} \hat{j}} \lambda_{\hat{k}}$}. Finally the commutator of the frames will be denoted as~${[\bh_a, \bh_b] = c^c_{~ab} \bh_c}$.

\section{Riemann--Cartan geometries}\label{sec:BasicRCG}

Let $M$ be a four-dimensional (4D) smooth manifold with coordinates $x^\mu$ with a non-degenerate coframe field $\bh^a$ defined on a subset $U \subset M$. We assume $M$ is equipped with a symmetric metric field $g_{ab}$ derived from the coframe fields to determine lengths and angles. Lastly, we require a~notion of parallel transport and assume the existence of a linear affine connection one-form~$\bomega^a_{~b}$. From the perspective of metric-affine theories, the geometrical quantities $g_{ab}$, $\bh^a$ and~$ \bomega^{a}_{~b}$ are independent quantities with 10, 16, and 64 independent elements, respectively \cite{Hehl_McCrea_Mielke_Neeman1995}. We will assume that our geometrical framework is invariant under the group of linear transformations of the frame ${\rm GL}(4,\mathbb{R})$. Further, we shall assume that the connection is compatible with the metric. These two assumptions impose constraints on the derived geometrical quantities.

\subsection{Gauge choices}

A choice of gauge will refer to the choice of basis for the tangent space. Using the ${\rm GL}(4, \mathbb{R})$ gauge, we can use the freedom in the tangent space to diagonalize the symmetric metric ${g_{ab}= \eta_{ab} = {\rm Diag}[-1,1,1,1]}$ with respect to the coframe $\{ \bh^a \} = ( {\bf u}, {\bf x}, {\bf y}, {\bf z})$; this is known as the \emph{orthonormal gauge} \cite{greub2012linear}. In this paper, $\{ \bh^a\}$ will always denote a frame basis in the orthonormal gauge. While other gauges are possible, such as the Newman--Penrose gauge \cite{kramer}, due to the natural $3+1$ split of spherically symmetric geometries, we choose to work in the orthonormal gauge.

There exists a subgroup of ${\rm GL}(4,\mathbb{R})$ of residual gauge transformations that leave the form of the metric unchanged. This subgroup is the orthogonal group~${\rm O}(1,3)$ under various representations. Due to physical motivations, we will require that the orientation of space and the direction of time cannot be changed. This constraint restricts the remaining gauge transformation from this subgroup to be the proper orthochronous Lorentz subgroup, ${\rm SO}^*(1,3)$ of ${\rm O}(1,3)$. We will denote the action of the Lorentz group on the coframe as \smash{$ \bh^a \to \Lambda^a_{~b} {\bh'}^b$}.

In the orthonormal gauge, the metric is completely fixed and only the coframe and connection are independent dynamical variables yielding $16+64=80$ independent elements with a~remaining~6-dimensional ${\rm SO}^*(1,3)$ gauge freedom. To obtain a Riemann--Cartan geometry, we impose the assumption that the connection be metric compatible, i.e., $Q_{abc}\equiv - \nabla_{c} g_{ab} =0$, whence the connection becomes anti-symmetric, $\bomega_{(ab)}=0$. Due to the algebraic nature of this constraint, it can be implemented easily without loss of generality.

The fundamental variables remaining are the 16 elements of the coframe $\bh^a$ and the 24 elements of the anti-symmetric connection one-form $\bomega^a_{~b}$. In terms of these quantities, the torsion tensor is of the form:
$T^a_{~bc} = \omega^a_{~cb} - \omega^a_{~bc} - c^a_{~bc}$,
 where $[\bh_c, \bh_d] = c^e_{~cd} \bh_e$ and $c^e_{~cd}$ are the coefficients of anholonomy for the frame $\bh_a$. The torsion tensor can be decomposed into three irreducible parts under the local Lorentz group \cite{Hehl_McCrea_Mielke_Neeman1995},
\[
T_{abc} = \frac23 (t_{abc} - t_{acb}) - \frac13 (g_{ab} V_c - g_{ac} V_b) + \epsilon_{abcd} A^d.
\]
 Here ${\bf V}$ denotes the vector part which is the trace of the torsion tensor:
$ V_a = T^b_{~ba}$.
 Lowering the index of the torsion tensor and applying the Hodge dual of the resulting tensor gives the axial part~${\bf A}$:
$A^a = \frac16 \epsilon^{abcd}T_{bcd}$.
 Finally, we can construct the purely tensorial part~${\bf t}$:
\[
 t_{abc} = \frac12 (T_{abc}+ T_{bac}) -\frac16 (g_{ca} V_b + g_{cb} V_a) + \frac13 g_{ab} V_c.
 \]
 We will call each of these tensors the {\it vector part, axial part, and tensor part} of the torsion tensor. The tensor part satisfies the following identities:
\[
 g^{ab} t_{abc} = 0, \qquad t_{abc} = t_{bac},\qquad t_{abc} + t_{bca} + t_{cab} = 0.
 \]

In a similar manner, the curvature tensor associated with the coframe and connection is defined as
\[
R^a_{~bcd} = \bh_c [\omega^a_{~bd}] - \bh_d [ \omega^a_{~bc}]+ \omega^e_{~bd} \omega^a_{~ec} - \omega^e_{~bc}\omega^a_{~ed}- c^e_{~cd} \omega^a_{~be},
\]
 where $[\bh_c, \bh_d] = c^e_{~cd} \bh_e$ and $c^e_{~cd}$ are, again, the coefficients of anholonomy for the frame $\bh_a$. This tensor can be decomposed into irreducible tensors \cite{obukhov2023poincare}: the Weyl tensor, two symmetric rank two tensors, an anti-symmetric rank two tensor, the Ricci scalar and a pseudo-scalar.
Through contractions of the Riemann tensor and its Hodge dual we may construct the Ricci tensor and co-Ricci tensor, respectively,
\[
	 R_{ab} = R^e_{~aeb}, \qquad
	 \bar{R}_{ab} = \frac12 R_{eacd} \epsilon^{cde}_{~~~b},
\]
 where $\epsilon^{abcd}$ is the Levi-Civita tensor associated with the Hodge dual. The Ricci scalar and pseudo-scalar follow as contractions of these tensors:
$R = R^a_{~a}$, $
	\bar{R} = \bar{R}^a_{~a}$.
 Denoting ${_\mathfrak{s}}T_{ab} = T_{(ab)}$ and ${_\mathfrak{a}} T_{ab} = T_{[ab]}$, respectively, as the symmetric and anti-symmetric parts of an arbitrary rank two tensor, ${\bf T}$, we may construct the trace-free Ricci tensor and trace-free co-Ricci tensor,
\[
	S_{ab} = {_\mathfrak{s}} R_{ab}-\frac14 R g_{ab}, \qquad
	\bar{S}_{ab} = {_\mathfrak{s}} \bar{R}_{ab}-\frac14 \bar{R} g_{ab}.
\]
 We note that ${_\mathfrak{a}} \bar{R}_{[ab]} = \frac12 \epsilon_{ab}^{~~cd} {_\mathfrak{a}} R$, and therefore we will only consider the anti-symmetric part of the Ricci tensor.

The irreducible parts of the curvature tensor are then \cite{obukhov2023poincare}
\begin{alignat*}{4}
		&{^{(2)}} R_{ab}^{~~cd} = \bar{S}_{e}^{~[c}\eta^{d]e}_{~~~ab}, \qquad&&
		{^{(3)}} R_{ab}^{~~cd} = - \frac{1}{12} \bar{R} \eta_{ab}^{~~cd}, \qquad&&
		{^{(4)}} R_{ab}^{~~cd} = - 2 S_{[a}^{~~[c} \delta^{d]}_{~~b]}, &\\
		&{^{(5)}} R_{ab}^{~~cd} = - 2{_\mathfrak{a}} R_{[a}^{~~[c} \delta^{d]}_{~~b]}, \qquad&&
		{^{(6)}} R_{ab}^{~~cd} = - \frac{1}{6} R \delta_{[a}^{~~[c} \delta^{d]}_{~~b]}.&&&
\end{alignat*}
The remaining irreducible part, the Weyl tensor, may be found by subtracting the above parts:%
\[
 {^{(1)}} R_{ab}^{~~cd} = C_{ab}^{~~cd} = R_{ab}^{~~cd} -   \sum_{I=2}^6 {^{(I)}} R_{ab}^{~~cd}.
 \]
 In this paper, we will work primarily with the Weyl tensor, the Ricci tensor, the torsion tensor and the covariant derivative of the vector part of the torsion tensor. Thus, we will ignore the finer decomposition of the Riemann tensor.

\section[The Cartan--Karlhede algorithm and the dimension of symmetry groups]{The Cartan--Karlhede algorithm\\ and the dimension of symmetry groups} \label{sec:CKalg}

The Cartan--Karlhede (CK) algorithm iteratively fixes a frame in an invariant manner by choosing canonical forms for the curvature tensor, the torsion tensor and their covariant derivatives. At the conclusion of the algorithm, the components of these tensors will be invariants for the geometry. However, these invariants only provide a local characterization of the geometry, and hence regularity within an open region is assumed so that the metric and its derivatives are continuous \cite[Chapter 9]{kramer}. This is necessary to avoid regions where the linear isotropy group drops in dimension, for example in the Szekeres Swiss-cheese models which arise from matching the Szekeres quasi-spherical solutions with spherically symmetric solutions, where the dimension of the linear isotropy group can drop from one to zero \cite{bolejko2010szekeres}.

To write the algorithm compactly, we will write $\mathcal{R}^q$ and $\mathcal{T}^q$ to denote the set of components of the curvature tensor and its covariant derivatives up to the $q$-th order, and the set of components of the torsion tensor and its covariant derivatives up to the $q$-th order, respectively.

The CK algorithm is then \cite{Coley:2019zld, fonseca1996algebraic, fonseca1992equivalence}:
\begin{enumerate}\itemsep=0pt
\item[(1)] Set the order of differentiation $q$ to 0.
\item[(2)] Calculate $\mathcal{R}^q$ and $\mathcal{T}^q$ .
\item[(3)] Determine the canonical form of the $q$-th covariant derivative of the curvature tensor and the torsion tensor.
\item[(4)] Fix the frame as much as possible, using this canonical form, and record the remaining frame transformations that preserve this canonical form (the group of allowed frame transformations is the {\it linear isotropy group $H_q$}).
\item[(5)] Find the number $t_q$ of independent functions of coordinates $x^\mu$ in $\mathcal{T}^q$ and $\mathcal{R}^q$ in the canonical form.
\item[(6)] If the dimension of $H_q$ and number of independent functions are the same as in the previous step, let $p+1=q$, and the algorithm terminates; if they differ (or if $q=0$), increase $q$ by~1 and go to step 2.
\end{enumerate}

If $\dim(H_p) \neq 0$, the frame resulting from the CK algorithm is known as an {\it invariantly defined frame fixed completely up to linear isotropy}. While if $H_p$ is the trivial group, so that $\dim(H_p) =0$, we say that the frame is an {\it invariant frame}. Relative to the frame determined by the algorithm, the non-zero components of $\mathcal{R}^q$ and $\mathcal{T}^q$ are then the set of {\it Cartan invariants} and we will denote the sets as $\mathcal{R} \equiv \mathcal{R}^{p+1}$ and $\mathcal{T} \equiv \mathcal{T}^{p+1}$.

In addition to these invariants, there are two sequences of discrete invariants: the dimension of the linear isotropy group, $\dim H_{q}$ at each iteration~$q$, and the number of functionally independent components of the curvature tensor, torsion tensor and their covariant derivatives up to order~$q$, denoted as~$t_q$. The linear isotropy group $H_q$ is a subgroup of the Lorentz frame transformations that do not change the form of the curvature tensor, the torsion tensor and their covariant derivatives up to $q$-th order. Assuming the frame basis and connection are sufficiently smooth, the 4D geometry is uniquely locally characterized by these two discrete sequences and the values of the (non-zero) Cartan invariants up to order $q$ which provides classifying functions.

The CK algorithm provides a straightforward approach to determining the number of symmetries of a Riemann--Cartan geometry. Since there are $t_p$ functionally independent invariants, they can be treated as essential coordinates~$x^{\alpha'}$. Then, the remaining $4-t_p$ coordinates $x^{\tilde{\alpha}}$ are ignorable, and so the dimension of the affine frame symmetry isotropy group (hereafter called the isotropy group) of the spacetime will be $s=\dim(H_p)$ and the affine frame symmetry group has dimension: $ r=s+4-t_p$.
Using this formula, we can determine the dimension for the largest group of affine frame symmetries by determining the best values for $t_p$ and $s$. Due to the negative sign in front of $t_p$, it follows that the best value is $t_p=0$ so that all Cartan invariants are constants. This is equivalent to the spacetime being locally homogeneous. We are left to consider the permitted values of $s$, which is determined by the linear isotropy of the Riemann curvature tensor and the torsion tensor. While the largest isotropy group for the Riemann curvature tensor is 6-dimensional, i.e., when the Weyl tensor vanishes and the Ricci tensor is proportional to the metric. However, the linear isotropy group of the torsion tensor was determined \cite{Coley:2019zld} and it was shown that the largest linear isotropy group is 3-dimensional. This occurs when the tensor part of the torsion tensor vanishes and both the vector part and axial part of the torsion tensor are proportional. Thus, the largest group of affine symmetries for Riemann--Cartan geometries with non-vanishing torsion is 7-dimensional.

\subsection{Lorentz frame transformations}
Any Lorentz frame transformation can be constructed in an orthonormal frame by combining boosts in planes spanned by the timelike direction and a spacelike direction, with spatial rotations in planes spanned by two spacelike directions.

We will briefly describe boost and rotations using two examples. For an orthonormal frame, if we consider a boost in the ${\bf u} - {\bf x}$ plane, with real-valued parameter $B$, the resulting transformation is then
\begin{gather} {\bf u}' = \frac{B^2+1}{2B} {\bf u} + \frac{B^2-1}{2B} {\bf x},\qquad
{\bf x}' = \frac{B^2-1}{2B} {\bf u} + \frac{B^2+1}{2B} {\bf x},\qquad
 {\bf y} ' = {\bf y},\qquad {\bf z}' = {\bf z}.\label{Ortho_Boost}
\end{gather}
 Similarly, a rotation in the ${\bf y}-{\bf z}$ plane, with parameter $\Theta$ is then
\begin{gather*}
{\bf u'} = {\bf u},\qquad {\bf x}' = {\bf x'}, \qquad
 {\bf y}' = \cos(\Theta) {\bf y} + \sin(\Theta) {\bf z}, \qquad
 {\bf z}' = -\sin(\Theta) {\bf y} + \cos(\Theta) {\bf z}.
\end{gather*}

\section{Affine frame symmetries with isotropy} \label{sec:FSs}

To discuss symmetries for a given Riemann--Cartan geometry $(M, {\bf g}, \bomega)$ in a frame context, we will consider the frame generated at the conclusion of the CK algorithm. The linear isotropy group $H_p$ encodes the freedom in specifying an invariantly defined (co)frame. That is, the CK algorithm produces an invariantly defined (co)frame $\bh^a$ up to linear isotropy $H_p$ such that for any diffeomorphism $\phi\colon M \to M$ that acts as an isometry of the metric, the invariantly defined (co)frame satisfies:
$ \phi^{*} \bh^a = \Lambda^a_{~b} \bh^b $,
 where $\Lambda^a_{~b} $ belongs to $H_p$ and arises from the coordinate form of~$\phi$, i.e., $y^\mu = \phi^\mu(x^\nu)$ \cite{olver1995equivalence}. If~${\phi = \phi_\tau}$ is generated by exponentiating the vector field which locally represents an infinitesimal generator of an affine frame symmetry ${\bf X}$ with some parameter~$\tau$, then we can calculate the effect of the Lie derivative of ${\bf X}$ on the frame as
\beq \mathcal{L}_{{\bf X}} \bh^a = \lambda^a_{~b} \bh^b, \label{Liederivative:frame} \eeq
 where $\lambda^a_{~b}$ is the Lie algebra generator for $\Lambda^a_{(\tau) b}$ associated with $\phi_{\tau}$. By using an invariantly defined frame, up to the linear isotropy group, we have reduced the number of unknown functions in $\lambda^a_{~b}$ and restricted their functional dependence to only those coordinates which are affected by $\phi_{\tau}$. Recall, to distinguish between those coordinates which are affected or unaffected by the affine frame symmetry generators we employ $x^{\alpha'}$ and $x^{\tilde{\alpha}}$, respectively.

In addition to the frame, we must consider the effect of the affine frame symmetry on the metric-compatible connection one-form $\bomega^a_{~b} = \omega^a_{~bc} \bh^c$, where $\bomega_{ab} = - \bomega_{ba}$. If ${\bf X}$ is an infinitesimal generator of an affine frame symmetry for the geometry, then it must also be an affine collineation~\cite{aaman1998riemann},
\beq (\mathcal{L}_{{\bf X}} \bomega)^a_{~bc} =0. \label{Liederivative:Con} \eeq

From the conditions \eqref{Liederivative:frame}--\eqref{Liederivative:Con}, it follows that the Lie derivative of the curvature tensor, the torsion tensor and its covariant derivatives with respect to ${\bf X}$ will all vanish. That is, any vector field satisfying equations \eqref{Liederivative:frame}--\eqref{Liederivative:Con} will necessarily satisfy equation \eqref{NaiveKV}. Similarly, the condition of a symmetry \eqref{NaiveKV} implies conditions \eqref{Liederivative:frame}--\eqref{Liederivative:Con} once an invariantly defined frame has been chosen. Hence this frame based definition of an affine frame symmetry is necessary and sufficient to generate an affine frame symmetry of the Riemann--Cartan geometry. In light of this, we introduce the following definition

\begin{Definition} \label{defn:AFS1}
An {\it affine frame symmetry} generator is a vector-field satisfying equations \eqref{Liederivative:frame} and~\eqref{Liederivative:Con}.
\end{Definition}

Note that if $\lambda^a_{~b} = 0$, then the original definition of an affine frame symmetry in \cite{Coley:2019zld} is recovered. That is, the linear isotropy group is trivial and the frame resulting from the CK algorithm is an invariant frame. In practice, when generating solutions, the above definition is not practical, as one requires knowledge of the curvature tensor, the torsion tensor and their covariant derivatives in order to determine the parameters of the Lorentz transformations.

We will instead consider a more general class of frames. These frames are acted on by the Lie algebra generators of the isotropy group under Lie differentiation with the affine frame symmetry generators.
\begin{Definition} \label{defn:SymFrame}
Consider a Riemann--Cartan geometry $({\bh}^a, \bomega^a_{~b})$ admitting an affine frame symmetry~${\bf X}$. The class of \emph{symmetry frames} $\bh^a$ are those frames that satisfy
\beq \mathcal{L}_{{\bf X}} \bh^a = f_{X}^{~\hat{i}} \lambda^a_{\hat{i}~b} \bh^b, \label{TP:frm:sym} \eeq
 where \smash{$f_X^{~\hat{i}}(x^{\alpha'})$} are arbitrary functions and \smash{$\lambda^a_{\hat{i}~b}$} are basis elements of the Lie algebra of the isotropy group (so that $\lambda_{\hat{i} ab} =- \lambda_{\hat{i} ba}$) and $\hat{i}$ ranges from 1 to the dimension of the isotropy group.
\end{Definition}

The functions \smash{$f_X^{~\hat{i}}$} have been introduced to permit the possibility of an arbitrary infinitesimal generator of an element of the isotropy group. This definition represents a choice of the frame basis and can be applied for any affine frame symmetry. We note that this definition has an advantage over the previous definition since $\Lambda^a_{~b}$ now has a definite structure that allows for the frame to be further fixed and determine the functional forms of the \smash{$f_{X}^{~\hat{i}}$} as well.

In general, a symmetry frame is not an invariantly defined frame until the components $f_{ X}^{~\hat{i}}$ are fixed in some coordinate independent way. A straightforward calculation shows that if ${\bf X}$ generates an affine frame symmetry, so that equation \eqref{TP:frm:sym} is satisfied, then the Lie derivative of the metric is zero, implying that ${\bf X}$ is a Killing vector field \cite{McNutt:2023nxmnxm}. Alternatively, given a Killing vector-field ${\bf X}$, then relative to an invariantly defined frame ${\bf X}$ satisfies equation \eqref{Liederivative:frame}, where~$\lambda^a_{~b}$ is a generator of an element of the linear isotropy group. As the linear isotropy group is a~subgroup of the larger isotropy group, this implies that $\bh^a$ is a symmetry frame and ${\bf X}$ is an affine frame symmetry.

Choosing a representation for the Lie algebra of the isotropy group, the transformation from the invariantly defined frame to an arbitrary symmetry frame can be stated in terms of the components of $f_X^{~\hat{i}}$. Since these components appear in equation~\eqref{TP:frm:sym}, they are tensor quantities that depend on the frame. Under a change of frame \smash{$\tilde{\bh}^a = \tilde{\Lambda}^a_{~b} \bh^b$} the components transform~as\looseness=1
\beq {\bf X}_I \big( \tilde{\Lambda}^a_{~b}\big) \big[\tilde{\Lambda}^{-1}\big]^b_{~c} + \tilde{\Lambda}^a_{~b} f_I^{~\hat{i}} \lambda^b_{\hat{i}~d} \big[\tilde{\Lambda}^{-1}\big]^d_{~c} = \tilde{f}_I^{~\hat{i}} \lambda^a_{\hat{i}~c}. \label{frot} \eeq
 Thus by starting with the invariantly defined frame, we may pick \smash{$\tilde{\Lambda}^a_{~b}$} to transform to any other symmetry frame by specifying the form of $\tilde{f}_I^{~\hat{i}}$. This shows that the class of symmetry frames always exists.

Given two different symmetry frames we can transform from one to the other using a subgroup of the Lorentz group $\overline{{\rm Iso}}$, which leaves the representation of the isotropy group unchanged. This can be determined using the chosen Lie algebra basis for the Lorentz group and picking those Lie algebra generators whose commutator with the Lie algebra elements of the isotropy group again lies in the isotropy algebra. For example, in the case of spherical symmetry there are three generators of the isotropy group~${\bf X}_I$, $I=1,2,3$, and the group $\overline{{\rm Iso}}$ consists of all spatial rotations.

In analogy with the CK algorithm, we fix the parameters of the elements in $\overline{{\rm Iso}}$ to prescribe the form of the $f_{ X}^{~\hat{i}}$ as much as possible. For example, some of the $f_{ X}^{~\hat{i}}$ can be set to a constant or zero. In which case, we have restricted the frame freedom of $\overline{{\rm Iso}}$ to a smaller subgroup $\overline{H}_p$, which leaves the chosen form of \smash{$f^{~\hat{i}}_{X}$} unchanged.

We will say that the resulting symmetry frame is an {\it invariantly defined symmetry frame up to the isotropy group $\overline{H}_p$}. While if $\overline{H}_p$ is the trivial group, the symmetry frame is an {\it invariant symmetry frame}. Returning to the spherically symmetric example, a choice of \smash{$f^{~\hat{i}}_{{X}_I} \equiv f^{~\hat{i}}_I$}, ${I=1,2,3}$, for each affine frame symmetry generator can be made so that this subgroup~$\overline{H}_q$ consists of rotations in the $\bh^3-\bh^4$ plane where their Lorentz parameters are constant.

Since the invariantly defined frame determined by the CK algorithm is a symmetry frame, it follows from equation \eqref{frot} that the original definition of an affine frame symmetry generator can be restated as

\begin{Definition} \label{AFS2}
An {\it affine frame symmetry} generator is a vector field satisfying
\begin{gather}
 \mathcal{L}_{{\bf X}_I} \bh^a = f_I^{~\hat{i}} \lambda^a_{~\hat{i}~b} \bh^b, \label{TP:sym1} \\
 (\mathcal{L}_{{\bf X}_I} \bomega)^a_{~bc} = 0 , \label{TP:sym2}
 \end{gather}
where $\hat{i}, \hat{j}, \hat{k} \in \{1, \dots, n\}$ and the components of $f_I^{~\hat{i}}$ are functions of the coordinates~$x^{\alpha'}$.
\end{Definition}

\subsection{Determining the most general geometry for a given isometry group} \label{subsec:FSgroups}

Supposing that the manifold admits a Lie group of affine frame symmetries of dimension $N$, with a non-trivial isotropy group with dimension $n$ $(n < N)$. Choose coordinates so that the affine frame symmetry group is represented as a set of vector fields~${\bf X}_I$, where $I, J, K \in \{1, \dots, N\}$, with the corresponding Lie algebra
\beq [{\bf X}_I, {\bf X}_J] = C^K_{~IJ} {\bf X}_K, \label{CijkSym} \eeq
 where $C^K_{~IJ}$ are structure constants of the Lie algebra.

Then we can determine the most general Riemann--Cartan geometries admitting this Lie group as affine symmetries by solving the following equations for an orthonormal symmetry coframe~$\bh^a$, the associated metric~\smash{$g_{\mu \nu} = \eta_{ab} h^a_{~\mu} h^b_{~\nu}$}, and with a metric-compatible connection~$\omega^a_{~bc}$ satisfying the conditions in definition~\eqref{AFS2}. The form of~$f_I^{~\hat{i}}$ can be specified by choosing some components to be equal to others, to be constant or to be zero. This defines a class of invariantly defined symmetry frames up to the linear isotropy in $\overline{H}_q$. For the moment, we will consider an arbitrary symmetry frame. In addition, due to the properties of the Lie derivative, we have an additional relationship coming from \eqref{CijkSym}:
\beq [\mathcal{L}_{{\bf X}_I}, \mathcal{L}_{{\bf X}_J}] \bh^a = \mathcal{L}_{[{\bf X}_I, {\bf X}_J] } \bh^a = C^K_{~IJ} \mathcal{L}_{{\bf X}_K} \bh^a. \label{TP:sym3} \eeq

Equations \eqref{TP:sym1}, \eqref{CijkSym} and \eqref{TP:sym3} determine the \emph{most general} frame basis which admits a given affine frame symmetry group with a non-trivial isotropy group. To determine the connection, we must solve equation \eqref{TP:sym2} relative to this frame. Using the coordinate basis expression for the Lie derivative of the connection in \cite{yano2020theory}, the corresponding frame basis expression is
\beq (\mathcal{L}_{{\bf X}_I}\omega)^a_{~bc} \bh_a = {\bf R}( {\bf X},\bh_c) \bh_b - {\bf T}({\bf X}, \nabla_c \bh_b) + \nabla_c {\bf T}({\bf X}, \bh_b) + \nabla_c \nabla_b {\bf X} - \nabla_{\nabla_c \bh_b} {\bf X}, \label{eq:RC:AC} \eeq
 where ${\bf R}(\bh_d, \bh_c)\bh_b = R^a_{~bcd} \bh_a$ and ${\bf T} ( \bh_c, \bh_b) = T^a_{~bc} \bh_a$, respectively, denote the curvature tensor and torsion tensor of the connection $\omega^a_{~bc}$.

Then denoting the associated Lie algebra of the isotropy group as
\[
[\lambda_{\hat{i}}, \lambda_{\hat{j}}]^a_{~b} = \lambda^a_{\hat{i}~b} \lambda^b_{\hat{j}~c} - \lambda^a_{\hat{j}~b} \lambda^b_{\hat{i}~c} = C^{\hat{k}}_{~\hat{i} \hat{j}} \lambda^a_{\hat{k}~c},
\]
 equations \eqref{TP:sym1}, \eqref{TP:sym3} and \eqref{eq:RC:AC} can be restated in the following proposition.

\begin{Theorem} \label{Sym:RC:Prop}
The most general Riemann--Cartan geometry admitting a given group of symmetries having vector generators, ${\bf X}_I$, $I,J,K \in \{1, \dots, N\}$ with a non-trivial isotropy group of dimension $n$ can be determined by solving for the unknowns $h^a_{~\mu}$, \smash{$f_I^{~\hat{i}}$} $($with $\hat{i}, \hat{j}, \hat{k} \in \{1,\dots, n\})$ and~$\omega^a_{~bc}$ from the following equations:
\begin{gather}
 { X}_I^{~\nu} \partial_{\nu} h^a_{~\mu} + \partial_{\mu} { X}_I^{~\nu} h^a_{~\nu} = f_I^{~\hat{i}} \lambda^a_{\hat{i}~b} h^b_{~\mu}, \label{TP:sym1big} \\
2{\bf X}_{[I} v( f_{J]}^{~\hat{k}}\big) - f_I^{~\hat{i}} f_J^{~\hat{j}} C^{\hat{k}}_{~~\hat{i} \hat{j}} = C^K_{~IJ} f_K^{~\hat{k}}, \label{TP:sym3big}\\
 { X}_I^{~d} \bh_d( \omega^a_{~bc}) + \omega^d_{~bc} f_I^{~\hat{i}} \lambda^a_{\hat{i}~d} - \omega^a_{~dc} f_I^{~\hat{i}} \lambda^d_{\hat{i} ~ b} - \omega^a_{~bd} f_I^{~\hat{i}} \lambda^d_{\hat{i}~c} - \bh_c\big( f_I^{~\hat{i}}\big) \lambda^a_{\hat{i}~b} = 0, \label{SFcon}
 \end{gather}
 where \smash{$\{ \lambda^a_{\hat{i}~b}\}_{\hat{i}=1}^n$} is a basis of the Lie algebra of the isotropy group, \smash{$[\lambda_{\hat{i}}, \lambda_{\hat{j}}] = C^{\hat{k}}_{~\hat{i}\hat{j}} \lambda_{\hat{k}}$}, $[{\bf X}_I, {\bf X}_J] = C^K_{~IJ} {\bf X}_K$.
\end{Theorem}

Given two symmetry frames, $\bh^a$ and $\tilde{\bh}^a$, the components \smash{$f_I^{~\hat{i}}$} in one frame can be related to~\smash{$\tilde{f}_I^{~\hat{i}}$} in the other frame using equation \eqref{frot} with the frame transformation $\tilde{\bh}^a = \tilde{\Lambda}^a_{~b} \bh^b$ determined by appropriately combining the respective frame transformations taking the symmetry frames to the same invariantly defined frame for the Riemann--Cartan geometry. While the above proposition applies for any symmetry frame, if one finds the solution in a particular symmetry frame, then the solution can be transformed to any other symmetry frame. In practice, the solution to these equations is most readily accomplished using an invariantly defined symmetry frame up to $\overline{H}_q$ where the components of $f_I^{~\hat{i}}$ have been invariantly specified. This is achieved by fixing the parameters of elements in the group $\overline{{\rm Iso}}$ to set the form of the matrix \smash{$f_I^{~\hat{i}}$} in an invariant manner.

These results can be applied to teleparallel geometries, by requiring that the connection is flat and so the curvature tensor expressed in terms of the connection and its derivatives must vanish.

\begin{Theorem}
The most general teleparallel geometry which admits a given group of symmetries with vector generators ${\bf X}_I$, $I,J,K \in \{1, \dots, N\}$ with a non-trivial isotropy group of dimension~$n$ can be determined by solving for the unknowns $h^a_{~\mu}$, \smash{$f_I^{~\hat{i}}$} $($with $\hat{i}, \hat{j}, \hat{k} \in \{1, \dots n\})$ and $\omega^a_{~bc}$ from Theorem~{\rm \ref{Sym:RC:Prop}} along with an additional condition
\[
 R^a_{\phantom{a}bcd} = h_c^{~\mu} \partial_\mu \omega^a_{\phantom{a}bd}- h_d^{~\nu}\partial_\nu \omega^a_{\phantom{a}bc}+\omega^a_{\phantom{a}fc}\omega^f_{\phantom{a}bd}-\omega^a_{\phantom{a}fd}\omega^f_{\phantom{a}bc} - c^e_{~cd} \omega^a_{~be} = 0,
\]
 where $[\bh_c, \bh_d] = c^e_{~cd} \bh_e$.
\end{Theorem}

\section[Riemann--Cartan geometries with SO(3)-isotropy]{Riemann--Cartan geometries with $\boldsymbol{{\rm SO}(3)}$-isotropy} \label{sec:3isospaces}

To illustrate the affine frame symmetry approach, we will review the construction of all Riemann--Cartan geometries admitting ${\rm SO}(3)$ isotropy \cite{McNutt:2023nxmnxm}. To do this, we will work in spherical coordinates $ \{x^{\mu}\} = ( t, r, \theta, \phi)$ so that the generators of the affine frame symmetry group are:
\[
{\bf X}_1 = \sin(\phi) \partial_\theta + \frac{\cos(\phi)}{\tan(\theta)} \partial_\phi,\qquad {\bf X}_2 = - \cos(\phi) \partial_\theta + \frac{\sin(\phi)}{\tan(\theta)} \partial_\phi,\qquad {\bf X}_3 = \partial_\phi.
\]
 The commutator relations for these vector fields are
$
 [{\bf X}_1, {\bf X}_2] = -{\bf X}_3$, $[{\bf X}_1, {\bf X}_3] = {\bf X}_2$ and $ [{\bf X}_2, {\bf X}_3] = -{\bf X}_1$.

To employ equation \eqref{TP:sym1big}, we must specify a representation of our isotropy subalgebra; we will choose the following:
\[
\lambda_{\hat{1}} = \left[ \begin{matrix} 0 & 0 & \hphantom{-}0 & 0\\ 0 & 0 & \hphantom{-}0 & 0 \\ 0 & 0 & \hphantom{-}0 & 1 \\ 0 & 0 & -1 & 0 \end{matrix}\right],\qquad
 \lambda_{\hat{2}} = - \left[ \begin{matrix} 0 & \hphantom{-}0 & 0 & 0\\ 0 & \hphantom{-}0 & 1 & 0 \\ 0 & -1 & 0 & 0 \\ 0 & \hphantom{-}0 & 0 & 0 \end{matrix}\right],\qquad
 \lambda_{\hat{3}} = - \left[ \begin{matrix} 0 & \hphantom{-}0 & 0 & 0\\ 0 & \hphantom{-}0 & 0 & 1 \\ 0 & \hphantom{-}0 & 0 & 0 \\ 0 & -1 & 0 & 0 \end{matrix}\right].
 \]
 With these quantities, we will determine a symmetry frame by specifying the form of $f_I^{~\hat{i}}$ in equation \eqref{TP:sym1big} using equation \eqref{TP:sym3big}. By exploiting the ${\rm SO}^*(1,3)$ freedom, these equations can be solved to give
\[
 f_I^{~\hat{i}} = \left[ \begin{matrix} \frac{\cos(\phi)}{\sin(\theta)} & 0 & 0 \vspace{1mm}\\
\frac{\sin(\phi)}{\sin(\theta)} & 0 & 0 \\
0 & 0 & 0 \end{matrix} \right] .
\]
With $f_I^{~\hat{i}}$ computed, we may solve equation \eqref{TP:sym1big} and determine the frame basis
\[
 \bh^1 = \alpha(t,r) {\rm d}t,\qquad \bh^2 = \beta(t,r) {\rm d}r,\qquad \bh^3 = \gamma(t,r) {\rm d}\theta,\qquad \bh^4 = \gamma(t,r) \sin(\theta) {\rm d}\phi.
 \]
 The functions $\a$, $\b$, $\e$ are arbitrary functions of $t$ and $r$. The coordinate freedom will be restricted to coordinate transformations that preserve the diagonal form of the metric.

Relative to this frame, we may then solve for the connection from equation \eqref{SFcon},
\begin{alignat*}{3}
& \bomega_{12} = W_5(t,r) \bh^1 + W_6(t,r) \bh^2, \qquad&&
\bomega_{13} = W_7(t,r) \bh^3 + W_8(t,r) \bh^4, &\\
&\bomega_{14} = - W_8(t,r) \bh^3 + W_7(t,r) \bh^3,\qquad&&
\bomega_{23} = W_3(t,r) \bh^3 + W_4(t,r) \bh^4, &\\
&\bomega_{24} = - W_4(t,r) \bh^3 + W_3(t,r) \bh^4,\qquad&&
\bomega_{34} = W_1(t,r) \bh^1 + W_2(t,r) \bh^2 - \frac{\cot(\theta)}{\e(t,r)} \bh^4,&
 \end{alignat*}
 where $W_a$, $a\in \{1, \dots, 8\}$, are arbitrary functions of $t$ and $r$. Of course, by imposing conditions on the torsion tensor or curvature tensor, and solving the resulting differential equations, we can recover solutions to Riemannian geometry if the torsion tensor vanishes or teleparallel geometry if the curvature tensor vanishes. Within Riemann--Cartan geometry, we can consider the existence of additional symmetries.

\section{Extension to static spherically symmetric geometries} \label{sec:G4spaces}

A popular choice for an additional symmetry is staticity, where a globally defined timelike affine frame symmetry exists and the resulting metric is diagonal in the same coordinate system where the timelike affine frame symmetry has been rectified. Without loss of generality, we may choose coordinates so that this timelike vector field takes the form:
${\bf X}_4 = \partial_t$,
 which does not affect the diagonal nature of the vielbein matrix $h^a_{~\mu}$. Then, the commutator relations immediately imply
$ [{\bf X}_{I'}, {\bf X}_4] = 0$, $ I' = 1,2,3$.
 By considering \smash{$f_I^{~\hat{i}}$}, where the index $I$ now ranges from 1 to 4, the frame freedom allows equation \eqref{TP:sym3big} to be solved again, giving
\[
 f_I^{~\hat{i}} = \left[ \begin{matrix} \frac{\cos(\phi)}{\sin(\theta)} & 0 & 0 \vspace{1mm}\\
\frac{\sin(\phi)}{\sin(\theta)} & 0 & 0 \\
0 & 0 & 0 \\
0 & 0 & 0 \end{matrix} \right].
\]
 Then solving equations \eqref{TP:sym1big} and \eqref{SFcon}, we find
\[
\bh^1 = \alpha(r) {\rm d}t,\qquad \bh^2 = \beta(r) {\rm d}r,\qquad \bh^3 = \gamma(r) {\rm d}\theta,\qquad \bh^4 = \gamma(r) \sin(\theta) {\rm d}\phi.
 \]

In general, in the static spherically symmetric Riemann--Cartan geometries, if $\frac{{\rm d}\e}{{\rm d}r} \neq 0$, then we may choose coordinates so that $\e(r) =r $ and the frame basis is now
\beq
\bh^1 = \alpha(r) {\rm d}t,\qquad \bh^2 = \beta(r) {\rm d}r,\qquad \bh^3 = r {\rm d}\theta,\qquad \bh^4 = r \sin(\theta) {\rm d}\phi, \label{StatSS:VB}
\eeq
and the connection coefficients then take the form
\begin{alignat}{3}
& \bomega_{12} = W_5(r) \bh^1 + W_6(r) \bh^2, \qquad&&
\bomega_{13} = W_7(r) \bh^3 + W_8(r) \bh^4, &\nonumber\\
&\bomega_{14} = - W_8(r) \bh^3 + W_7(r) \bh^3, \qquad&&
\bomega_{23} = W_3(r) \bh^3 + W_4(r) \bh^4, &\nonumber\\
&\bomega_{24} = - W_4(r) \bh^3 + W_3(r) \bh^4, \qquad&&
\bomega_{34} = W_1(r) \bh^1 + W_2(r) \bh^2 - \frac{\cos(\theta)}{r \sin(\theta)} \bh^4.& \label{SO3connection}
\end{alignat}

\section[Static Spherically Symmetric G\_7 Riemann--Cartan geometries]{Static Spherically Symmetric $\boldsymbol{G_7}$ Riemann--Cartan geometries} \label{sec:SO3 Case 2}

In principle, additional symmetries can be appended to the group. However, this requires knowledge of the larger Lie algebra that contains the Lie subalgebra constructed from the current affine frame symmetry generators. Furthermore, not all Lie algebras can be realized as affine frame symmetry generators for a given class of Riemann--Cartan geometries. A good example of this phenomenon is the de Sitter analogue in Riemann--Cartan geometry, which is a~spherically symmetric geometry admitting a seven-dimensional group of affine frame symmetries. The non-trivial de Sitter group's ten-dimensional Lie algebra is not realized in Riemann--Cartan geometry.

In the case of large affine frame symmetry groups, there is a constructive approach to determining the geometries that admit them. We will use the tools introduced above to determine a subclass of static spherically symmetric Riemann--Cartan geometries that admit seven affine frame symmetry generators, $G_7$.

From the CK algorithm, a Riemann--Cartan geometry will admit such a group of affine frame symmetries if $\dim H_p = 3$ and the components of the torsion tensor, the curvature tensor and their covariant derivatives are constant relative to an invariantly defined frame; that is, $t_p = 0$. As an illustration, in the following section we will outline the analysis for the subclasses of static spherically symmetric Riemann--Cartan geometries and determine the conditions in each inequivalent case. The results of this analysis will be gathered in Appendix \ref{apA}.

Using the frame basis in equation \eqref{StatSS:VB}, along with the connection in equation \eqref{SO3connection}, we can compute the curvature tensor and torsion tensor. The tensor part of the torsion tensor admits at most a two-dimensional linear isotropy group \cite{Coley:2019zld}. In order to realize a three-dimensional linear isotropy group, this tensor~${\bf t}$ must necessarily vanish. The vanishing of the tensor part of the torsion tensor gives
\[
 W_4 = W_2,\qquad W_8=W_1,\qquad W_5 = W_3 + \frac{1}{\b r}- \frac{\a_{,r}}{\a \b},\qquad W_7 = W_6.
 \]
 The axial and vector parts of the torsion tensor are then
\[{\bf A} = 2 W_2 \bh^1 + 2 W_1 \bh^2, \qquad
{\bf V} = -3W_6 \bh^1 - 3\left(W_3 - \frac{1}{r \b} \right) \bh^2.
 \]

The components of the torsion tensor and curvature tensor are not necessarily constant in the current frame. We will consider a boosted frame~$\{ \bh_a\}_{a=1}^4$, where
\[
 {\bf A} = 2C_1 {\bh^1}', \qquad {\bf V} = -3C_2 {\bh^1}'
 \]
 for some real-valued constants $C_1$ and $C_2$. We will consider the one-form
 \[
 {\bf \tilde{V}} = {\bh^1}' = (2C_1)^{-1} {\bf A} = (-3C_2)^{-1} {\bf V}.
 \]
 The boost parameter in equation \eqref{Ortho_Boost} can be written as $B = {\rm e}^{f(r)}$, and we can rewrite the remaining components of the connection as
\begin{gather*}
 W_1 = C_1 \sinh(f), \qquad
W_2 = C_1 \cosh(f),\qquad
 W_3 =-\frac{1}{r \b}+ C_2 \sinh(f),\\
 W_6 = C_2 \cosh(f).
\end{gather*}
 We will continue to work with respect to the boosted frame to determine conditions on the remaining free functions using the covariant derivatives of $\bh_1'$ and the Ricci tensor and the Weyl tensor. In the case of the Weyl tensor, it necessarily must vanish as the largest linear isotropy group for a non-vanishing Weyl tensor is at most 2-dimensional \cite{kramer}.

In order for the Ricci tensor to admit a 3-dimensional linear isotropy group, the following component must vanish:
\beq R_{12} = \frac{{\rm e}^{-2f}({\rm e}^{4f} -1)(- \a \b_{,r} - \b \a_{,r})}{2\a \b^3 r}. \label{StaticR12}\eeq
 This vanishes if the boost parameter is unity (i.e., if $f = 0$) or if the vielbein functions satisfy the following constraint, $D_0 = \a \b$.

Recall, when we started the analysis for this section we assumed $\frac{{\rm d}\e}{{\rm d}r} \neq 0 $ and chose coordinates so that $\gamma = r$. If we were to repeat the analysis here assuming $\gamma = \gamma_0$ is constant, then the equivalent of equation~\eqref{StaticR12} implies that $\a$ or $\b$ must vanish, and so it is not permitted.
\begin{enumerate}\itemsep=0pt
\item[(1)] $\a\b \neq \text{ constant}$ and $f = 0$:
If $\nabla_a \tilde{V}_b$ admits a 3-dimensional linear isotropy group, then $\nabla_2 \tilde{V}_1 =0$ and we find the following condition
$-\frac{2}{\b} (\ln \a)_{,r} = 0$
which implies $\a = D_1$, and this may be rescaled so that~${D_1=1}$, giving
\smash{$ \nabla {\bf \tv} = 2 C_2 \big({\bf g} - \big|{\bf \tv}\big|^{-2} {\bf \tv} {\bf \tv}\big)$}.
 Imposing the same condition on the Ricci tensor, we find that
\smash{$ \b = \pm \frac{1}{\sqrt{D_2 r^2 +1}}$},
 with the corresponding Ricci tensor and co-Ricci tensor, respectively,
\begin{gather*}
	 {\bf R} = \bigl(-2C_1^2 +2 C_2^2 - 2 D_2\bigr) \big({\bf g} - \big|{\bf \tv}\big|^{-2} {\bf \tv} {\bf \tv}\big), \\
	 {\bf \bar{R}} = (2 C_1 C_2) \big({\bf g} - \big|{\bf \tv}\big|^{-2} {\bf \tv} {\bf \tv}\big) + 6 C_1 C_2 \big|{\bf \tv}\big|^{-2} {\bf \tv} {\bf \tv}.
\end{gather*}
 The Weyl tensor vanishes automatically and all other components of the Riemann tensor are constants. The coframe and connection that gives these $G_7$ geometries are explicitly given in Appendix \ref{apA}.

\item[(2)] $\a \b = D_0$ and $f = 0$:
We may rescale the arbitrary functions $\a$ and $\b$ so that $D_0 = 1$. This is a special case of the next case where $f = 0$. We will omit the analysis for this case and note that the coframe and connection that gives these $G_7$ geometries are explicitly given in Appendix~\ref{apA}, case 2 where $f=0$.

\item[(3)] $\a \b = D_0$:
Again, we may rescale the arbitrary functions $\a$ and $\b$ so that $D_0 = 1$. By integrating the numerator of the component $\nabla_1 \tilde{V}_2 = 0$ with respect to $r$, we find that
$ - 2\a \cosh(f) = D_1$.
The vanishing of $\nabla_1 \tilde{V}_2$ in combination with the equation $1 = \a \b$ implies that the vierbein functions are
\[
 \a = -\frac{D_1}{2\cosh(f)},\qquad \b =- \frac{2\cosh(f)}{D_1},\qquad \e = r.
 \]
Requiring that $\nabla_2 \tilde{V}_2 = \nabla_3 \tilde{V}_3 = C_3$ gives
\[
 2C_2 - \frac{4 {\rm e}^{2f} D_1 f_{,r}}{\big({\rm e}^{2f} +1\big)^2} = C_3, \qquad
 2 C_2 - \frac{D_1 \sinh(f)}{r \cosh(f)} = C_3.
 \]
 Note that $C_3 = 2C_2$ is not permitted as this would imply $D_1 =0 $ and hence $\a =0$, which cannot happen. Solving for $f$ gives
\[
 f = \frac12 \ln \left( -\frac{r(2C_2 - C_3) + D_1 }{r(2C_2 - C_3) - D_1} \right).
 \]

The Ricci tensor admits a 3-dimensional linear isotropy group if $D_1 = \pm 2$, while the Weyl tensor automatically vanishes. Thus, the non-vanishing rank two tensors are
 \begin{gather*}
 \nabla {\bf \tv} = C_3 \big({\bf g} - \big|{\bf \tv}\big|^{-2} {\bf \tv} {\bf \tv}\big), \\
{\bf R} = \left[-2C_1^2-\frac14 C_3(2C_2 - 3C_3)\right]\big({\bf g} - \big|{\bf \tv}\big|^{-2} {\bf \tv} {\bf \tv}\big) - \frac{3}{4} (2C_2 - C_3) C_3 \big|{\bf \tv}\big|^{-2} {\bf \tv} {\bf \tv},\\
{\bf \bar{R}} = (-2C_1(C_2-C_3))\big({\bf g} - \big|{\bf \tv}\big|^{-2} {\bf \tv} {\bf \tv}\big) + 3C_1C_3 \big|{\bf \tv}\big|^{-2} {\bf \tv} {\bf \tv}.
\end{gather*}
 The coframe and connection that yields these $G_7$ geometries are explicitly given in Appendix \ref{apA}.
\end{enumerate}

\section[Stationary spherically symmetric Riemann--Cartan geometries and their G\_7 subclasses]{Stationary spherically symmetric Riemann--Cartan\\ geometries and their $\boldsymbol{G_7}$ subclasses} \label{sec:SO3 case 1}

If we consider spherically symmetric Riemann--Cartan geometries that are stationary but not static, a straightforward computation shows that
$ {\bf X}_4 = -C_0 \partial_t + r \partial_r$,
 with $C_0 \neq 0$. As in the static case, we find $[{\bf X}_{I'}, {\bf X}_4] =0$ for $I' =1,2,3$. Due to this commutator structure, the analysis in Section \ref{sec:G4spaces} may be repeated to determine the coframe and the connection.

In the original coordinates, the coframe functions take the form
\[
 \a(t,r) = \a\big(r{\rm e}^{\frac{t}{C_0}}\big), \qquad\b(t,r) = \b\big(r{\rm e}^{\frac{t}{C_0}}\big){\rm e}^{\frac{t}{C_0}},\qquad \e(t,r) = \e\big(r{\rm e}^{\frac{t}{C_0}}\big) .
 \]
 We may choose new coordinates $\{ t', r', \theta, \phi\}$ so that ${\bf X}_4 = \partial_{t'}$. In this frame, the coframe basis is then
\begin{gather}
 \bh^1 = \a(r') {\rm d}t',\qquad \bh^2 = -r' \b(r') {\rm d}t' + \b(r') {\rm d}r',\qquad \bh^3 = \e(r') {\rm d}\theta,\nonumber\\
 \bh^4 = \e(r') \sin \theta {\rm d}\phi, \label{Coframe Stationary}
 \end{gather}
 from which it follows that the vielbein matrix is no longer diagonal in this coordinate system.
 In this coordinate system the connection one-forms are now
\begin{alignat}{3}
& \bomega_{12} = W_5(r') \bh^1 + W_6(r') \bh^2, \qquad&&
\bomega_{13} = W_7(r') \bh^3 + W_8(r') \bh^4,&\nonumber \\
&\bomega_{14} = - W_8(r') \bh^3 + W_7(r') \bh^3, \qquad&&
\bomega_{23} = W_3(r') \bh^3 + W_4(r') \bh^4, &\nonumber\\
&\bomega_{24} = - W_4(r') \bh^3 + W_3(r') \bh^4, \qquad&&
\bomega_{34} = W_1(r') \bh^1 + W_2(r') \bh^2 - \frac{\cos(\theta)}{\e(r')\sin(\theta)} \bh^4.&\label{SO3connection Stationary}
\end{alignat}

\subsection[G\_7 Stationary spherically symmetric Riemann--Cartan geometries]{$\boldsymbol{G_7}$ Stationary spherically symmetric Riemann--Cartan geometries}

In what follows, we will drop the prime on the coordinates and work in the coordinate system where ${\bf X}_4$ has been rectified, so that ${\bf X}_4 = \partial_{t}$. With the coframe and connection given in equations \eqref{Coframe Stationary} and \eqref{SO3connection Stationary}, we can compute the curvature tensor and torsion tensor. We ask that the tensor-part of the torsion vanishes in order to ensure that the linear isotropy group of the torsion tensor is three-dimensional. This gives the following conditions:
\[
W_4 = W_2,\qquad W_8=W_1,\qquad W_5 = W_3 + \frac{\e_{,r}}{\b \e}- \frac{\a_{,r}}{\a \b},\qquad W_7 = W_6 - \frac{\e_{,r} r}{\a \e} + \frac{\b_{,r} r}{\a \b}+\frac{1}{\a}.
 \]

Using these conditions, the axial and vector parts of the torsion become
\[
 {\bf A} = 2 W_2 \bh^1 + 2 W_1 \bh^2, \qquad
{\bf V} = -3 \left( W_6 + \frac{\b_{,r} r}{\a \b}
+\frac{1}{\a} \right) \bh^1 - 3 \left( W_3 + \frac{\e_{,r}}{\b \e}\right) \bh^2.
\]
 There is a new frame $\{\bh_a'\}_{a=1}^4$ arising from applying a boost in the $\bh_1 - \bh_2$ plane, where ${{\bf A} = 2C_1 {\bh^1}'}$ and ${\bf V} = -3C_2 {\bh^1}'$ for some real-valued constants $C_1$ and $C_2$. We find that for the boost parameter in equation \eqref{Ortho_Boost}, $B = {\rm e}^{f(r)}$, we have additional conditions on the components of the connection,
\begin{gather*}
 W_1 = C_1 \sinh(f(r)), \qquad
 W_2 = C_1 \cosh(f(r)), \qquad W_3 = -\frac{\e_{,r}}{\b \e} + C_2 \sinh(f(r)), \\
 W_6 = - \frac{r \b_{,r}}{\a \b} - \frac{1}{\a} + C_2 \cosh(f(r)).
\end{gather*}
 We will now work with respect to the boosted frame and determine conditions on the remaining free functions using the Ricci tensor and the covariant derivative of ${\bf \tilde{V}}$, where
 \[{\bf \tilde{V}} = {\bh^1}' = (2C_1)^{-1} {\bf A} = (-3C_2)^{-1} {\bf V}. \]

Computing the Ricci tensor, we find the following component must vanish if it is to admit a~3-dimensional linear isotropy group
\[
 R_{12} = -\frac{{\rm e}^{-2f}}{2 \b^3 \a^3 \e} \big({\rm e}^{4f} (r\b -\a)^2 - (r\b+\a)^2 \big)(\b \a_{,r} \e_{,r} - \b \a \e_{,r,r} + \a \b_{,r} \e_{,r}).
 \]
 This vanishes if $f = \frac14 \ln \bigl( \frac{(r \b+\a)^2}{(r \b-\a)^2} \bigr)$ or $\e_{,r} = \alpha \beta D_0^{-1}$.

We will first consider the case, where \smash{$f = \frac14 \ln \bigl( \frac{(r \b+\a)^2}{(r \b-\a)^2}\bigr)$}, and we will consider the components of $\nabla_a \tilde{V}_b$. In order for $\nabla_a \tilde{V}_b$ to admit a 3-dimensional linear isotropy group, the following must vanish:
\[
 \nabla_3 \tilde{V}_3 - \nabla_2 \tilde{V}_2 = \frac{\e \big(\b^2 r^2 - \a^2\big)_{,r} - 2 \e_{,r} \big(r^2 \b^2 - \a^2\big) }{\a \b \e \sqrt{r^2 \b^2 - \a^2}} = 0.
 \]
 This leads to the following solution $\e^2 = D_1 \big(\b^2 r^2 - \a^2\big)$. We will choose the following para\-metrization for the functions
\[
 \a = \sqrt{|D_1|} \e \sinh(g(r)),\qquad \b = \frac{ \sqrt{|D_1|} \e \cosh(g(r))}{r},
 \]
 From the condition on $f$, it follows that $f = g$. However, by computing the Weyl tensor, there are components of this tensor of the form $\e^{-1}$ and hence can never vanish. Thus, this tensor can never admit a 3-dimensional linear isotropy group and there are no $G_7$ solutions in this case.

The only possibility of attaining $G_7$ stationary spherically symmetric Riemann--Cartan geometries are if the coframe functions satisfy:
\smash{$ \e_{,r} = \frac{\a \b}{D_0}$}. We will rescale the arbitrary functions~$\a$ and $\b$ so that $D_0 = 1$. To continue, we will consider the components of $\nabla_a \tilde{V}_b$ and $R_{ab}$ relative to the boosted frame to get further conditions on the remaining functions.

Starting with the component $\nabla_2 \tilde{V}_1$, setting this to zero and integrating the numerator with respect to $r$ gives
\beq 2r \b \sinh(f) - 2\a \cosh(f) = D_1, \label{SO3_split_c1a} \eeq
 where $D_1$ is an arbitrary constant of integration. Here, we can consider two subcases, when~${f \neq 0}$ and when $f = 0$.
\begin{enumerate}\itemsep=0pt
\item[(1)] $f=0$:
 If $f=0$, then we can write $\a = D_1$. Examining the remaining non-zero components in~$\nabla_a \tilde{V}_b$, we have
\[
 \nabla_2 \tilde{V}_2 = \frac{2(C_2 D_1 \b - \b_{,r} r -\b)}{ D_1 \b}, \qquad
\nabla_3 \tilde{V}_3 = \nabla_4 \tilde{V}_4 = \frac{2C_2(\e-r \b)}{\e}.
 \]
 These components should be constants and equal to the same value $ \nabla_2 \tilde{V}_2 = \nabla_3 \tilde{V}_3 = C_3$.
 The second equation $\nabla_3 \tilde{V}_3 = C_3 $ requires that $C_3 \neq 2 C_2$ as this would imply that $\b$ would vanish, which cannot occur. Solving the first solution for $\b$, we find
\[
\b = D_2 r^{\frac{D_1(2C_2 - C_3)-2}{2}},\qquad C_3 \neq 2 C_2.
 \]
Solving for $\e$ from this equation gives
\[
\e = \frac{2 D_2 r^{\frac{D_1(2C_2 - C_3)}{2}}}{2C_2 - C_3}.
\]

By rescaling the $r$ coordinate, we can set $D_2 = 1$. We will examine the components of the Ricci tensor. In particular, we ask that $R_{22} - R_{33} = 0$ giving an algebraic condition on $D_1$
\[
 \frac{D_1^2-1}{4} (2C_2- C_3)^2 r^{-D_1 (2C_2 - C_3)} = 0.
 \]

Imposing that $D_1 = \pm 1$, the Ricci tensor now admits a 3-dimensional linear isotropy group%
\[
 {\bf R} = \left[-2C_1^2-\frac14 C_3 (2C_2 - 3 C_3)\right] \big({\bf g} - \big|{\bf \tv}\big|^{-2} {\bf \tv} {\bf \tv}\big) - \frac{3}{4} C_3 (2C_2 - C_3) \big|{\bf \tv}\big|^{-2} {\bf \tv} {\bf \tv}.
 \]

 All components of the Weyl tensor vanish automatically. The Ricci tensor and co-Ricci tensor take the form
\begin{gather*}
 {\bf R} = \left[-2C_1^2-\frac14 C_3 (2C_2 - 3 C_3) \right] \big({\bf g} - \big|{\bf \tv}\big|^{-2} {\bf \tv} {\bf \tv}\big) - \frac{3}{4} C_3 (2C_2 - C_3) \big|{\bf \tv}\big|^{-2} {\bf \tv} {\bf \tv}, \\
{\bf \bar{R}} = -2 C_1 (C_2 - C_3) \big({\bf g} - \big|{\bf \tv}\big|^{-2} {\bf \tv} {\bf \tv}\big) + 3 C_1 C_3 \big|{\bf \tv}\big|^{-2} {\bf \tv} {\bf \tv} .
\end{gather*}
 In Appendix \ref{apB}, we summarize the coframe and connection that gives rise to the $G_7$ stationary spherically symmetric Riemann--Cartan geometries listed below.

\item[(2)] $f \neq 0$: If $f \neq 0$, the algebraic condition in equation \eqref{SO3_split_c1a} implies that $\a$ and $\b$ take the form
\[
 \a = \frac{D_1 g(r)}{2 \cosh(f)},\qquad \b = \frac{D_1 (1-g(r))}{2 r \sinh(f)}.
 \]
 The remaining non-zero components satisfy $\nabla_i \tilde{V}_i = C_3$ for $i=2,3,4$, this is now an algebraic equation for $g(r)$ with the solution
\[
 g = \frac{- \e \sinh(2f) (C_3-2C_2) }{2 D_1} + \frac{1 + \cosh(2f)}{2}.
 \]
 Substituting this into the differential equation for $\e$, $\e_{,r} = \a \b$, we can algebraically solve for the boost parameter
\[
 {\rm e}^{2f} = \frac{8 \e_{,r} r \pm \sqrt{\e^4 (C_3 - 2C_2)^4 - 2 D_1^2 \e^2 (C_3 - 2C_2)^2 + 64 \e_{,r}^2 r^2 + D_1^4 }}{D_1^2 + \e^2 (C_3 - 2C_2)^2-4\e (C_3 - 2C_2)}.
\]
Using these expressions in the Ricci tensor, there is one additional condition that must be satisfied before the tensor admits a 3-dimensional linear isotropy group
\[
 R_{22} - R_{33} = \frac{D_1^2 - 4}{4\e^2} = 0.
 \]
 Thus, $D_1 = \pm 2$ and now the Ricci tensor and the co-Ricci tensor take the form
 \begin{gather*}
{\bf R} = \left[-2C_1^2 -\frac14 C_3 (2C_2 - 3 C_3)\right] \big({\bf g} - \big|{\bf \tv}\big|^{-2} {\bf \tv} {\bf \tv}\big) - \frac{3}{4} C_3 (2C_2 - C_3) \big|{\bf \tv}\big|^{-2} {\bf \tv} {\bf \tv}, \\
{\bf \bar{R}} = -2C_1(C_2-C_3) \big({\bf g} - \big|{\bf \tv}\big|^{-2} {\bf \tv} {\bf \tv}\big) + 3C_1 C_3 \big|{\bf \tv}\big|^{-2} {\bf \tv} {\bf \tv}.
 \end{gather*}
 The Weyl tensor vanishes automatically, implying that these solutions admit a $G_7$ group of symmetries.

\end{enumerate}

\section{Discussion} \label{sec:Discussion}

In this paper, we have discussed the development of symmetry techniques in Riemann--Cartan geometries. While it may be natural to consider a modification of the Killing equations in equation \eqref{NaiveKV} to determine affine frame symmetries, the study of symmetries are more easily developed using a frame formalism. In this perspective, the metric is replaced with a frame basis which diagonalizes the metric and the torsion tensor is replaced with the connection coefficients. Then the generator, ${\bf X}$, of an affine frame symmetry of the Riemann--Cartan geometry is an affine collineation of the space where the Lie derivative of the frame basis with respect to ${\bf X}$ is given by equation \eqref{Liederivative:frame}.

While these conditions may appear more complicated than the original formulation in terms of the metric and the torsion, there are two tools that have been developed to investigate symmetries. The first tool is the Cartan--Karlhede (CK) algorithm, which was originally developed to locally characterize a given Riemann--Cartan geometry in terms of invariants by determining an invariantly defined frame. A natural byproduct of this algorithm is a simple calculation to determine the number of affine frame symmetries in a given geometry. The second tool is the affine frame symmetry approach, which expands the class of invariantly defined frames to symmetry frames and allows for the explicit calculation of all Riemann--Cartan geometries which admit a desired affine frame symmetry group.

Using the symmetry frame approach, we have presented the construction of all spherically symmetric Riemann--Cartan geometries and identified the subclass of static spherically symmetric Riemann--Cartan geometries. Then by employing the CK algorithm, we determined conditions for a given static spherically symmetric geometry to admit a seven-dimensional affine frame symmetry group~$G_7$. By explicitly solving these conditions for the frame basis and the connection coefficients, we determined all inequivalent static spherically symmetric Riemann--Cartan $G_7$ geometries, which are summarized in Appendix~\ref{apA} below.

We note that this is one possible subclass of spherically symmetric $G_7$ Riemann--Cartan geometries, and in the pursuit of determining all spherically symmetric Riemann--Cartan geometries admitting a $G_7$ group of affine frame symmetries we examined the stationary spherically symmetric Riemann--Cartan geometries. These admit a timelike affine frame symmetry ${\bf X}_4$ but the metric cannot be diagonalized in the same coordinate system, where ${\bf X}_4$ has been rectified. Repeating the approach used for the static case, we determined all inequivalent subclasses that admit a $G_7$ group of affine frame symmetries which has been summarized in Appendix \ref{apB} below.

In this paper, we have determined all spherically symmetric Riemann--Cartan geometries that admit a $G_7$ group of affine frame symmetries. More generally, since the maximal linear isotropy group for the torsion tensor is at most three-dimensional, and can be realized as ${\rm SO}(3)$, ${\rm SO}(1,2)$ or ${\rm E}(2)$, we anticipate there are $G_7$ Riemann--Cartan geometries outside of the spherically symmetric class. This will be explored in future work.

\appendix

\section[Static spherically symmetric G\_7 Riemann--Cartan geometries]{Static spherically symmetric $\boldsymbol{G_7}$ Riemann--Cartan geometries}\label{apA}

The connection one-forms for any static spherically symmetric $G_7$ Riemann--Cartan are
 \begin{alignat*}{3}
& \bomega_{12} = W_5(r) \bh^1 + W_6(r) \bh^2, \qquad&&
\bomega_{13} = W_6(r) \bh^3 + W_1(r) \bh^4, &\\
&\bomega_{14} = - W_1(r) \bh^3 + W_6(r) \bh^3, \qquad&&
\bomega_{23} = W_3(r) \bh^3 + W_2(r) \bh^4, &\\
&\bomega_{24} = - W_2(r) \bh^3 + W_3(r) \bh^4, \qquad&&
\bomega_{34} = W_1(r) \bh^1 + W_2(r) \bh^2 - \frac{\cos(\theta)}{r \sin(\theta)} \bh^4.&
 \end{alignat*}
\begin{enumerate}\itemsep=0pt
\item[(1)] Case 1: $\a \b \neq$ constant and $f = 0$.
The frame basis takes the form
\[
 \bh^1 = {\rm d}t,\qquad \bh^2 = \pm \frac{1}{\sqrt{D_2 r^2 +1}} {\rm d}r,\qquad
\bh^3 = r {\rm d}\theta,\qquad \bh^4 = r \sin(\theta) {\rm d}\phi,
\]
 where the components of the connection one-forms are
\begin{gather*}
 W_1 = 0,\qquad W_2 = C_1,\qquad W_3 =\pm \frac{ \sqrt{D_2 r^2 +1}}{r},\qquad W_5 = \pm \frac{2 \sqrt{D_2 r^2 +1}}{r },\\ W_6 = C_2.
 \end{gather*}
\item[(2)] Case 2: $\a \b=1$ writing $k = \pm \frac{2C_2-C_3}{2}$, the frame basis is now
\begin{gather*}
 \bh^1 = \pm \sqrt{1-k^2r^2} {\rm d}t,\qquad\! \bh^2 = \pm \left(\! \frac{1}{\sqrt{1-k^2r^2} } \!\right) {\rm d}r,\qquad\! \bh^3 = r {\rm d}\theta,\qquad\! \bh^4 = r \sin(\theta) {\rm d}\phi
 \end{gather*}
 with boost parameter
 \[
 B = {\rm e}^{f(r)}\colon \
 f = \frac12 \ln \left( \frac{r k + 1}{1- r k} \right),
 \]
 where the connection one-form components are
\begin{gather*}
 W_1 = \frac{C_1 k r}{\sqrt{1-k^2r^2} },\qquad W_2 = \frac{C_1}{\sqrt{1-k^2r^2} } ,\qquad
 W_3 =\pm \frac{\sqrt{1-k^2r^2} }{r} + \frac{C_2 k r}{\sqrt{1-k^2 r^2} }, \\
 W_5 = \mp \frac{k^2r}{\sqrt{1- k^2 r^2}}- \frac{C_2 k r}{\sqrt{1-k^2r^2}}, \qquad
W_6 = \frac{C_2}{\sqrt{1-k^2r^2} }.
\end{gather*}
\end{enumerate}

\section[Stationary spherically symmetric G\_7 Riemann--Cartan geometries]{Stationary spherically symmetric\\ $\boldsymbol{G_7}$ Riemann--Cartan geometries}\label{apB}

Relative to the coordinate system, where ${\bf X}_4 = \partial_t$ defined in Section \ref{sec:SO3 case 1}, the coframe is then
\[
 \bh^1 = \a(r) {\rm d}t,\qquad \bh^2 = -r \b(r) {\rm d}t + \b(r) {\rm d}r,\qquad \bh^3 = \e(r) {\rm d}\theta,\qquad \bh^4 = \e(r) \sin \theta {\rm d}\phi,
 \]
 while the connection one-forms are
\begin{alignat*}{3}
& \bomega_{12} = W_5(r) \bh^1 + W_6(r) \bh^2, \qquad&&
\bomega_{13} = W_7(r) \bh^3 + W_1(r) \bh^4, &\\
&\bomega_{14} = - W_1(r) \bh^3 + W_7(r) \bh^3,\qquad&&
\bomega_{23} = W_3(r) \bh^3 + W_2(r) \bh^4, &\\
&\bomega_{24} = - W_2(r) \bh^3 + W_3(r) \bh^4, \qquad&&
\bomega_{34} = W_1(r) \bh^1 + W_2(r) \bh^2 - \frac{\cos(\theta)}{\e(r)\sin(\theta)} \bh^4.&
\end{alignat*}
\begin{enumerate}\itemsep=0pt
\item[(1)] Case 1: $f=0$.
The coframe functions takes the form
\[
\a = \pm 1,\qquad \b = r^{\frac{\pm (2C_2 -C_3)-2}{2}},\qquad \e = \frac{2 r^{\frac{\pm (2C_2 -C_3)}{2}}}{2C_2 - C_3},
\]
 where $C_3 \neq 2 C_2$. The arbitrary functions in the connection one forms are then
 \begin{gather*}
 W_1 = 0, W_2 = C_1,\qquad W_3 = -\frac12 (2C_2 - C_3)r^{-\frac12 (2C_2 - C_3)}, \qquad W_5 = 0,\\ W_6 = \frac12 C_3,\qquad W_7 = \frac12 C_3 .
\end{gather*}
\item[(2)] Case 2: $f \neq 0$.
The coframe functions takes the form
\[
 \a = \frac{\pm g(r)}{\cosh(f(r))},\qquad \b = \frac{\pm(1-g(r))}{r \sinh(f(r))},\qquad \e = \e(r) ,
 \]
where $\e$ is an arbitrary function of $r$ and $g$ is defined as
\[
 g = \frac{\pm \e \sinh(2f)(C_3-2C_2)}{4} + \frac{1+\cosh(2f)}{2}.
 \]

The boost parameter is then defined as
\[
 {\rm e}^{2f} = \frac{8 \e_{,r} r \pm \sqrt{\e^4 (C_3 - 2C_2)^4 - 8 \e^2 (C_3 - 2C_2)^2 + 64 \e_{,r}^2 r^2 + 16 }}{4 + \e^2 (C_3 - 2C_2)^2-4\e (C_3 - 2C_2)}.
 \]
The remaining functions of the connection one-forms are then
 \begin{gather*}
 W_1 = C_1 \sinh(f),\qquad W_2 = C_1 \cosh(f), \qquad W_3 = -\frac{\e_{,r}}{\b \e} + C_2 \sinh(f), \\
W_5 = C_2 \sinh(f) - \frac{\a_{,r}}{\a \b},\qquad W_6 = - \frac{r \b_{,r}}{\b \a} - \frac{1}{\a} + C_2 \cosh(f), \\ W_7 = C_2 \cosh(f) - \frac{\e_{,r} r}{\a \e}. \end{gather*}

\end{enumerate}

\subsection*{Acknowledgements}
The authors would like to thank the anonymous referees for their helpful comments which have improved the quality of the paper. AAC and RvdH are supported by the Natural Sciences and Engineering Research Council of Canada. RvdH is supported by the Dr.\ W.F.~James Chair of Studies in the Pure and Applied Sciences at St.\ Francis Xavier University. DDM is supported by the Norwegian Financial Mechanism 2014-2021 (project registration number 2019/34/H/ST1/00636).

\pdfbookmark[1]{References}{ref}
\LastPageEnding

\end{document}